\documentstyle[12pt]{article}

\def\Journal#1#2#3#4{{#1} {\bf #2}, #3 (#4)}

\def\NPB{{\em Nucl. Phys.} B}
\def\PLB{{\em Phys. Lett.}  B}
\def\PRL{\em Phys. Rev. Lett.}
\def\PRD{{\em Phys. Rev.} D}

\def\be{\begin{equation}}
\def\ee{\end{equation}}
\def\bea{\begin{eqnarray}}
\def\eea{\end{eqnarray}}
\newcommand{\k}{\vec{k}_{\perp}^2}
 \newcommand{\la}{\langle}
\newcommand{\ra}{\rangle}

\newcommand{\Dm}{\vec{iD}_{\mu} }

\begin{document}
\begin{titlepage}
\begin{flushright}
 hep-ph/9605226\\
May 1996
\end{flushright}
\vspace{0.3cm}

 \begin{center}
\Large\bf
Constituent Quark Model versus Nonperturbative QCD.
\end{center}
 \vspace {0.3cm}

 \begin{center}
 {\bf Ariel R. Zhitnitsky}\footnote{
  e-mail address:arz@physics.ubc.ca }
 \end{center}
\begin{center}
{\it Physics Department, University of British Columbia,
6224 Agricultural  Road, Vancouver, BC V6T 1Z1, Canada}
\end{center}
 \begin{abstract}

   We discuss a
 few, apparently different
(but actually, tightly related) problems: \\
1. The relation between QCD and valence quark model,\\
 2. The asymptotic behavior of  the  nonperturbative
  pion wave function $\psi(\k, x)$ at $x\rightarrow 0,~1, \k
\rightarrow \infty$  and  \\
3. The dimensional counting rules in the intermediate region of energy.
The analysis is
based on such general methods as    dispersion relations,
duality and  PCAC.
We calculate the asymptotic behavior of the
wave function ($wf$)
at the end-point region ($x\rightarrow 1$ and  $\k\rightarrow\infty$)
by analysing  the corresponding  large  $n-$th  moments  in transverse
$\la\vec{k}_{\perp}^{2n}\ra \sim n!$ and   longitudinal
$\la (2x-1)^n\ra \sim 1/n^2$ directions. This information  fixes  the
asymptotic behavior of $wf$ at large $\k$ ( which
 is turned    out to be Gaussian commonly used in the phenomenological
analyses).
 We discuss some applications of the obtained results.
In particular, we calculate the nonleading ``soft"
contribution to the pion form factor at intermediate momentum
transfer.   We argue, that due to the
specific properties of   $\psi(\k, x)$, the corresponding contribution
can  temporarily {\bf simulate} the leading twist behavior
in the extent region of $ Q^2 $.
The same effect   also takes place for the nucleon formfactor.
  Such a mechanism, if it is correct, would be
 an explanation of the phenomenological success of the dimensional
counting rules   at available, very modest energies
for many different processes. We discuss some inclusive amplitudes
 also.

\begin{center}
{\it Talk given at the Workshop ``Continuous Advances in QCD,96''.\\
Theoretical Physics Institute, University of Minnesota, \\Minneapolis,
MN, March 28-31, 1996.}
\end{center}
\end{abstract}
\end{titlepage}
\vskip 0.3cm
\noindent
\setcounter{page}{1}
 \section{Motivation. Definitions.}
  The problem of bound states in the
relativistic quantum field theory with
large coupling constant is an extremely difficult
problem. To  understand the structure of the
bound state is a very ambitious goal which
assumes the
solution of a whole spectrum of tightly connected problems, such as
confinement,
chiral symmetry breaking phenomenon,   and many others which
are greatly important in the low energy region.

A less ambitious purpose is the  study of
 the hadron wave function ($wf$) with a minimal number
of constituents\footnote{
 Such a study is the
  way to  understand   the valence quark model in   the QCD terms.
This remark explains  the title of this talk.}.
As is known such a function gives the  parametrically leading contributions
to hard exclusive processes.  The corresponding
wave functions within QCD  have  been introduced to the theory
 in the late seventies
and  early eighties  \cite{Brod}   to describe the
exclusive processes. We refer to the review papers
\cite{Cher}$^,$\cite{Brod1}
 on this subject for the  details.

The main idea of the approach \cite{Brod}$^-$\cite{Brod1}
 is the separation of the large and small
distance physics. At small distances we can use the standard
perturbative expansion due to the asymptotic freedom and
smallness of the coupling constant. All nontrivial,
large distance physics is hidden into the
nonperturbative wave function ($wf$)
and can not be found
by perturbative technique, but rather it should be extracted
from  elsewhere.   The most powerful analytical
nonperturbative method
for such   problems, I believe, is the
QCD sum rules \cite{Shif1}$^,$\cite{Shif2}.

The first application of QCD sum rules to the analysis
of nonperturbative $wf$ was considered more than decade ago
\cite{Cher2}.  The information   extracted
for the few lowest
moments, unambiguously shows the asymmetric form of the
distribution amplitudes.
At the same time, the applicability of the approach
\cite{Brod}  at experimentally
accessible momentum transfers was questioned  \cite{Isgur}.
 Since then this subject
  is a very controversial
issue     and we are not going
to comment  these quite opposite  points
in the present talk.  Instead we would like
  to formulate the following question:

$\bullet$  If the asymptotically leading contribution can not provide
the experimentally observable absolute
values, than {\it how   can one
explain the very good agreement between the
experimental data and  dimensional counting rules}
 which are supposed to be valid  only in the region where the
leading terms dominate?

It is clear, that the possible explanation can not be related to the
specific amplitude, but instead, it should be
connected, somehow, to the nonperturbative
wave functions  of the light hadrons ($\pi, \rho, N...$)
which enter  the formulae for exclusive processes.
The analysis of the $\pi$ meson and nucleon
form factors, presented below supports this idea.

To anticipate the events we would like  to formulate here the result
of this analysis.
The very unusual properties
  of a nonperturbative hadron wave function lead to the
{\bf temporarily simulation} of the dimensional counting rules
by soft  mechanism for the $F_{\pi}Q^2,~F_{N}Q^4$
in the extent range of intermediate momentum transfer:
$3 GeV^2\leq Q^2\leq 40 GeV^2 $. In this region   the soft contribution
 to the form factors does not fall-off, as
naively one could expect.

$\bullet$ Therefore, our answer on the formulated question
is the  following.  The
light cone nonperturbative wave functions
are very different from the   ones, motivated
by the naive quark model.
In the former, QCD case,
  there is a dimensional parameter
which determines all scales in the problem.
This parameter is nothing, but a mean value of quark transverse momentum
$\la\k\ra\simeq (330 MeV)^2$ which can be expressed in terms of
QCD vacuum condensates.
In the latter, valence quark model case,
there is one more parameter,   the  constituent quark mass $m$,
which has  the same  order of magnitude.
 In spite of the fact that both these scales are
numerically very close to each other, they are fundamentally different.
 The difference is due to the fact that
$m\simeq 330 MeV$ is a constant
while $0\leq\k\leq\infty$ is a variable
with average $\la\k\ra\simeq (330MeV)^2$.

 Let me emphasize from the
very beginning  that
 the ideology and methods (unitarity,
dispersion relations, duality) we use are motivated
by QCD sum rules.  However,  we do not use the  QCD sum rules
in the common sense: we do not fit them
  to extract any information about lowest resonance
(as  people usually  do in this approach), we do not
use any numerical approximation or implicit assumption about
higher states. Instead, we concentrate on analysis of the
appropriate correlation functions themselves   to extract
 the most general information.

The starting point is the  definition  of   $wf$
in terms of nonperturbative matrix elements.
To be more specific, let us consider  the $\pi-$ meson
axial wave function:
\bea
\label{d}
if_{\pi}q_{\mu}\phi_A (zq,z^2)=
\la 0|\bar{d}(z)\gamma_{\mu}\gamma_5
e^{ig\int_{-z}^z A_{\mu}dz_{\mu}} u(-z)|\pi(q)\ra   \\
\nonumber
 =\sum_n \frac{i^n}{n!}\la 0|\bar{d}(0)\gamma_{\mu}\gamma_5
(iz_{\nu}\stackrel{\leftrightarrow}{D_{\nu}})^n u(0)|\pi(q)\ra ,
\eea
where
$\stackrel{\leftrightarrow}{D_{\nu}}\equiv
\stackrel{\rightarrow}{D_{\nu}}-\stackrel{\leftarrow}{D_{\nu}}$ and
$\Dm=i\vec{\partial_{\mu}}+gA_{\mu}^a\frac{\lambda^a}{2}$ is the
covariant derivative.
{}From its definition is clear that the set of different
$\pi$ meson matrix elements defines the nonperturbative wave
function. Exactly this definition of the $wf$
as the set of different matrix elements we have in mind
when we discuss the nonperturbative $wf$.
Such a $wf$ may or may not satisfy some equations.

First of all, let us discuss,
the most important part (at  asymptotically high $q^2$)
of the $wf$ which is
  related to the longitudinal distribution.
In this case $z^2\simeq 0$  and the $wf$
depends on  $zq$- variable only. The corresponding
Fourier transformed
wave function will be denoted as $\phi(\xi)$
and its
  $n-$th moment is given by the
following local matrix element:
\be
\label{1}
\la 0|\bar{d}\gamma_{\nu}
\gamma_5(i\stackrel{\leftrightarrow}{D_{\mu}}
z_{\mu})^{n} u|\pi(q)\ra=if_{\pi}q_{\nu}
(zq)^{n} \la \xi^{n}\ra=if_{\pi}q_{\nu}(zq)^{n}
\int^1_{-1}d\xi\xi^{n}\phi(\xi)
\ee
\be
\label{}
-q^2\rightarrow\infty,~~zq\sim 1~~
\xi=x_1-x_2,~~ x_1+x_2=1,~~z^2=0.   \nonumber
\ee
 Therefore, if we knew all matrix elements (\ref{1})
( which are well-defined ) we could restore the whole
distribution amplitude $\phi(\xi)$.

 Now we would like to analyse  the similar  moments, but
in transverse direction. To do so, let us define the
transverse vector $t_{\mu}=(0,\vec{t},0)$ to be  perpendicular
  the hadron momentum $q_{\mu}=(q_0,0_{\perp},q_z)$.
The  vector $t_{\mu}$ is an appropriate projector in   transverse
plane and plays the same role what $z_{\mu}$ vector
does in eq.(\ref{1}) for longitudinal direction.
We define
  the $2n-th$ moment for the  transverse quark distribution
in the following way:
\be
\label{6}
\la 0|\bar{d}\gamma_{\nu}
\gamma_5 (\stackrel{\rightarrow}{iD_{\mu}}
 t_{\mu})^{2n} u|\pi(q)\ra=if_{\pi}q_{\nu}
 (-t^2)^n\frac{(2n-1)!!}{(2n)!!}\la \vec{k}_{\perp}^{2n} \ra.
\ee
where $\stackrel{\rightarrow}{D_{\nu}}$ is the
covariant derivative,
acting on the one quark.
 The factor $\frac{(2n-1)!!}{(2n)!!}$ is introduced   to
(\ref{6}) to take into account
  the integration over $\phi$ angle in the transverse plane:
$\int d\phi (\cos\phi)^{2n}/ \int d\phi= {(2n-1)!!}/{(2n)!!}$.

We interpret the $\la \vec{k}_{\perp}^{2} \ra$ in this equation
as  a mean value square of the transverse quark   momentum. Of course it
is different from the naive, gauge dependent  definition like
$\la 0|\bar{d}\gamma_{\nu}
\gamma_5 \partial_{\perp}^2 u|\pi(q)\ra $,
because the physical transverse gluon is participant
of this definition.
However, the expression (\ref{6}) is the only possible
way to define the  moments $\la \vec{k}_{\perp}^{2n} \ra$
in the gauge theory like QCD.
 We believe that such definition is the useful generalization
of the transverse momentum conception for the interactive quark
system.

The definition of the  nonperturbative wave function
is as follows: Let us assume that we know all longitudinal moments
(\ref{1}) as well as  transverse moments (\ref{6}).
We introduce the wave function of two variables
$\psi(\k, \xi )$, normalized to one
\be
\label{a6}
\int d\k \psi(\k, \xi )
= \phi(\xi),~\int_{-1}^1 d\xi\phi(\xi)=1
\ee
  in such a way that
it exactly reproduces the set of moments defined by the local
matrix elements (\ref{1},\ref{6}),
 The relations to Brodsky and Lepage
notations $\Psi_{BL}(x_1,\vec{k}_{\perp})$
\cite{Brod1}   looks as follows:
\be
\label{9}
\Psi_{BL}(x_1,\vec{k}_{\perp})=\frac{f_{\pi}16\pi^2
}{\sqrt{6}}\psi(\xi,\vec{k}_{\perp}),~
\int d\k d\xi\psi(\k , \xi )=1
\ee
where $f_{\pi}=133 MeV$.
\section{Constraints}
{}As me mentioned earlier, the QCD sum rules is
appropriate method to analyse the matrix elements
given by the formule (\ref{1},\ref{6}). We start our discussion
from the analysis of the longitudinal moments.

The corresponding calculations have been done
many years ago and  a few first moments have been estimated\cite{Cher2}.
However, this information is  not enough to reconstruct
the $wf$; the parametric behavior at $\xi\rightarrow\pm 1$
is the crucial issue in this reconstruction.

To extract the corresponding information,
we use the following duality argument.
Instead of consideration of the  pion $wf$ itself, we study
  the following correlation function with pion quantum numbers:
\be
\label{2}
i  \int dx e^{iqx}\la 0|T J_{n}^{\|}(x),J_0(0) |0\ra=
(zq)^{n+2}I_{n}(q^2),~~
 J_{n}^{\|}=\bar{d}\gamma_{\nu}z_{\nu}
\gamma_5(i\stackrel{\leftrightarrow}{D_{\mu}}z_{\mu})^{n} u
 \ee
and calculate its asymptotic behavior at large $q^2$.
The result can be presented in the form of the dispersion integral,
whose spectral density is determined by the pure perturbative one-loop
diagram:
\be
\label{3}
\frac{1}{\pi}\int_0^{\infty} ds\frac{Im I_n^{pert}(s)}{s-q^2},~~
 Im I_{n}(s)^{pert}=\frac{3}{4\pi(n+1)(n+3)}.
 \ee
  We {\bf assume } that the $\pi$ meson gives a
nonzero contribution to the dispersion integral for arbitrary
$n$ and, in particular, for $n\rightarrow\infty$.
Formally, it can be written in the following way
 \be
\label{3a}
 \frac{1}{\pi}\int_0^{S_{\pi}^n} ds Im I(s)^{pert}_{n}=
\frac{1}{\pi}\int_0^{\infty} ds Im I(s)^{\pi}_{n},
 \ee
   Our assumption means
   that $S_{\pi}^n(\|)\neq 0$,
where we specified the notation for the longitudinal distribution.
In this case at $q^2\rightarrow\infty$ our assumption (\ref{3a})
leads to the following relation:
\be
\label{4}
  f_{\pi}^2\la \xi^n\ra (n\rightarrow\infty)
\rightarrow\frac{3S_{\pi}^{\infty}(\|)}{4\pi^2n^2}
 \ee
It  unambiguously  implies the  following behavior at
the end-point region \cite{Cher}:
 \be
\label{5}
 \la\xi^n\ra=\int_{-1}^1d\xi \xi^n\phi(\xi)\sim 1/n^2,~~~~~
   \phi(\xi\rightarrow
\pm 1)\rightarrow (1-\xi^2).
\ee

 Thus, our first  constraint   looks as follows:

$\bullet 1~~~~~~~~~~~~~~~~~~~~~~~
   \phi(\xi\rightarrow
\pm 1)\rightarrow (1-\xi^2)$.

We want to emphasize that
we did not use any numerical approximation
in this derivation. Therefore, the constraint ($\bullet 1$)
 has very general origin
 and it should be considered
as a direct consequence of QCD. Only dispersion relations, duality
and very plausible assumption formulated above have been used in
the derivation ($\bullet 1$).
We can repeat these arguments for the analysis of the transverse
distribution.
The  result is\footnote{ Here
and in what follows we ignore any mild (nonfactorial) $n$-dependence.}:
\be
\label{8}
f_{\pi}^2\la \vec{k}_{\perp}^{2n} \ra \frac{(2n-1)!!}{(2n)!!}\sim
S_{\pi}^{n+1}(\perp)n!\Rightarrow
 f_{\pi}^2\la \vec{k}_{\perp}^{2n} \ra \sim n!,~~ n\rightarrow\infty .
\ee
 This behavior has been obtained in ref.\cite{Zhit2}
by analysing the perturbative series  of the specific
correlation function at large order. The dispersion relations
and duality arguments translate this information into the
formula (\ref{8}).

 The nice feature of   (\ref{8})
is its finiteness  for arbitrary $n$. It means
that the higher  moments
\be
\label{8a}
\la \vec{k}_{\perp}^{2n} \ra =
\int d\k d\xi
\vec{k}_{\perp}^{2n}\psi(\k, \xi )
\ee
{\bf do exist}.
The  existence of the arbitrary high moments
$\la \vec{k}_{\perp}^{2n} \ra$ means that the nonperturbative
$wf$, defined above, falls off at large transverse momentum $\k$
faster than any power function.
   The relation (\ref{8})
fixes the asymptotic behavior of $wf$ at large $\k$.
Thus, we arrive to the following constraint:

$\bullet 2~~~~~~~~~~~~~~~~
   \la \vec{k}_{\perp}^{2n} \ra =
\int d\k d \xi
\vec{k}_{\perp}^{2n}\psi(\k, \xi )\sim n!
 ~~~~~n\rightarrow\infty  .$

We can repeat our duality arguments again for an arbitrary number of
transverse derivatives and large ($n\rightarrow\infty$) number
of longitudinal derivatives with the following result \cite{Zhit1}:

$\bullet 3~~~~~~~~~~~~~~~~~~
  \int d\k
\vec{k}_{\perp}^{2k}\psi(\k, \xi\rightarrow\pm 1 )\sim  (1-\xi^2)^{k+1}.$

For   $k=0$ we reproduce our previous formula
for the $\phi$   function: $\phi (\xi\rightarrow\pm 1)=
\int d\k \psi(\k, \xi\rightarrow\pm 1)\sim (1-\xi^2) $.
The constraint ($\bullet 3$)
 is extremely important and implies that the $\k$
dependence of the   $\psi(\k, \xi )$ comes
{\bf exclusively in the combination}
$ \k/(1-\xi^2)$ at $\xi\rightarrow\pm 1$.  The byproduct of this
constraint can be formulated as follows. The standard assumption
on factorizability of the $\psi(\k, \xi ) =\psi(\k )\phi(\xi)$
{\bf does contradict} to the very general properties of the theory.
 Thus, the asymptotic behavior of the $wf$   turns out to be Gaussian one
with the very specific argument:
\be
\label{!}
 \psi(\k\rightarrow\infty, \xi)\sim \exp(-\frac{\k}{(1-\xi^2)} ),
 \ee
 Let us remark, that the same methods can be applied for the analysis
of the asymptotical behavior of the nucleon $wf$ as well.
In obvious notations the asymptotic behavior for the
nucleon $wf$ takes  the form:
\be
\label{nucl}
\psi_{nucleon}(\vec{k}_{\perp i}^2
\rightarrow\infty, x_i)\sim \exp(-\sum\frac{ \vec{k}_{\perp i}^2}{x_i}).
\ee
\section{QCD vs. valence Quark Model}
 We would like to   make the following conjecture:
The Gaussian $wf$ reconstructed earlier from the
QCD analysis (\ref{!}, \ref{nucl}),
 not accidentally coincides with the harmonic oscillator
$wf$  from the constituent quark model.
 To make this conjecture more clear,    let us recall
 few results  from the constituent quark model.

It is well known \cite {Isgur1} that the equal- time
 wave functions
\be
\label{qm}
\psi_{CM}(\vec{q}\,^2)\sim \exp (-\vec{q}\,^2)
\ee
of the harmonic oscillator in the rest frame
give a very reasonable description of
static meson properties.
Together with Brodsky-Huang-Lepage prescription
\cite{BHL}
connecting  the equal -time   and the
light-cone wave functions of two  constituents (with mass $m\sim  300 MeV$)
by identification
\be
\label{BHL}
  \vec{q}^2\leftrightarrow\frac{\k+m^2}{4x(1-x)}-m^2,
{}~~\psi_{CM}(\vec{q}^2)\leftrightarrow\psi_{LC}
(\frac{\k+m^2}{4x(1-x)}-m^2),
\ee
one can reproduce the
Gaussian behavior (\ref{!}) found  from QCD.
It means, first of all, that our identification
of the moments (\ref{6}) defined in  QCD
with the ones defined in quark model,
is the reasonable conjecture.

 However, there is a difference. In valence quark model
we do have a parameter which describes the mass
of constituents $m\simeq 330 MeV$. We have nothing like that
in QCD.    Indeed the presence of  such a    term in QCD,
    would mean
 the following behavior for the
large moments in the longitudinal direction:
   \be
\label{21}
 \la\xi^n\ra=\int_{-1}^1d\xi \xi^n\phi(\xi)
\sim \int_{-1}^1d\xi \xi^n\exp(-\frac{1}{1-\xi^2})\sim
\exp(-\sqrt{n}), ~n\rightarrow\infty.
  \ee
This is in contradiction with  $1/n^2$ behavior
(\ref{5}), ($\bullet 1$) found earlier.
 We do not see any possibilities to change this behavior
from $1/n^2$ to $ \exp(-\sqrt{n})$ within QCD. Thus, the massive term
in the $wf$ motivated by  valence quark model
has no any justification from the QCD point of view.
The same conclusion is also true for the nucleon $wf$.
Again, we have no room for the mass term in the formula
(\ref{nucl}). This observation, as we shall see has very
important impact on the phenomenological analysis.
   \section{Numerical Constraints.}

In the previous sections  we discussed
some general   $wf$  properties  which should be satisfied
for any QCD-based model.
However, those constraints do not determine
the scale of the problem; they do not give a
dimensional factor which would govern the hadronic
properties. In the present section we want to
discuss some  numerical ( and therefore, less general)
constraints on   wave functions.

 Thus, there is a big    difference
between  constraints
($\bullet 1-\bullet 3$) discussed above
and  the ones which follow. The first three constraints
have  very general origin. No numerical approximations have been made
in the derivation of the corresponding formulae.
The constraints which follow have
absolutely different status. They are based on the QCD sum rules
with their inevitable numerical assumptions about  higher
excited states in QCD. Thus, they must be treated as
an approximate ones. The well-known constraint of such
a kind is the second moment of the distribution
amplitude in  the
longitudinal direction \cite{Cher2}:

$\bullet 4 ~~~~~~~~~~~~~~~~~~~~~~
\la\xi^2\ra \equiv \int d\xi \phi(\xi)\xi^2\simeq 0.4,$

 (The asymptotic $wf$ corresponds to  $\la\xi^2\ra\ = 0.2$ ).
Such a result was the reason to suggest   the ``two-hump" shape $wf$
\cite{Cher2}
which meets the above  requirement. The number
cited as  the constraint ($\bullet 4$)
has been   criticized in the literature.
Thus, in what follows we shall discuss both possibilities: the narrow
(asymptotic)  $wf$  and  the wider one (with larger $\la\xi^2\ra\ >0.2$).

The next ``numerical" constraint
is the second moment of the $wf$
in the transverse direction defined by equation
(\ref{6}) and calculated for the  first time in \cite{Cher} and
independently (with quite different technique) in \cite{Novik}.
Both results are in a full agreement to each other:

$\bullet 5 ~~
 \la \vec{k}_{\perp}^{2} \ra=\frac{5}{36}\frac{\la
\bar{q}ig\sigma_{\mu\nu}
 G_{\mu\nu}^a\frac{\lambda^a}{2} q \ra }{\la \bar{q}q\ra}
\simeq \frac{5 m_0^2}{36}\simeq 0.1GeV^2, ~~~m_0^2\simeq 0.8 GeV^2.$

Essentially, the constraint ($\bullet 5$) defines the general scale
of all nonperturbative phenomena for the  pion. It is not
accidentally coincides with $ 330 MeV$   which is the
typical magnitude in the hadronic physics.

  To study  the fine properties of the
transverse distribution it is desired to know the next moment.
The problem can be reduced to the analysis of the
 mixed vacuum condensates of dimension seven \cite{Zhit1}:
 \be
\label{12}
  \la\vec{k}_{\perp}^{4}\ra =
 \frac{1}{8}\{   \frac{-3\la \bar{q}g^2\sigma_{\mu\nu}
 G_{\mu\nu} \sigma_{\lambda\sigma}
 G_{\lambda\sigma}q \ra}{4\la\bar{q}q\ra}
+ \frac{ 13\la \bar{q}g^2
 G_{\mu\nu}   G_{\mu\nu}q \ra}{ 9 \la\bar{q}q\ra}\},
\ee
We analyzed the magnitudes for  these vacuum condensates with
the following result: the standard factorization hypothesis does not work
in this case. The factor of nonfactorizability $K\simeq 3.0\div 3.5$
\cite{Zhit1}$^,$\cite{Zhit3}.
The eq.(\ref{12}) defines the new numerical constraint on the transverse
distribution.
We prefer to express this constraint not in  terms of the absolute
values, but rather, in terms of the dimensionless parameter $R$
which is defined in the following way:
$$\bullet6 ~~~~~~~~~ R\equiv
\frac{\la \vec{k}_{\perp}^{4}\ra  }{\la \vec{k}_{\perp}^{2}\ra^2}
\simeq 3K\cdot\frac{ \la g^2G_{\mu\nu}^aG_{\mu\nu}^a\ra}{m_0^4}
\simeq  5\div 7, ~~~~~~~~~~ m_0^2\simeq 0.8GeV^2    ,$$
 where we use the standard values for parameter $m_0^2$ and gluon
condensate \cite{Shif2}. We would like to emphasize
that the fluctuations of the transverse momentum are large enough. The   quantitative
 characteristic of these fluctuations is parameter
$R\gg1$. In terms of the wave function this property means a very
unhomogeneous distribution in transverse direction.
\section{Applications.}
 Having found the constraints on the $wf$ in the previous sections,
one can model them and finally, one can apply them
for the calculation of different amplitudes.
The corresponding calculations he been carried out
for the $\pi$ meson in ref.\cite{Chibisov}and for the
nucleon in ref.\cite{Kroll}.
We quote here some results of the corresponding calculations.

The first application is the pion form factor.
 The starting point is the   Drell-Yan formula
  where $F_{\pi}(Q^2)$ is expressed in terms
of the wave functions:
\be
\label{DY}
F_{\pi}(Q^2)=\int\frac{dx d^2\vec{k}_{\perp}}
{16\pi^3}\Psi^*_{BL}
(x ,\vec{k}_{\perp}+(1-x) \vec{q}_{\perp})
\Psi_{BL}(x ,\vec{k}_{\perp}),
\ee
where $q^2=-\vec{q}_{\perp}^2=-Q^2$ is the momentum transfer.
In this formula, the $\Psi_{BL}(x ,\vec{k}_{\perp})$ is
 the full wave function;
the perturbative tail of $\Psi_{BL}(x ,\vec{k}_{\perp})$ behaves as
$\alpha_s/ \k$ for large $\k$ and should be taken into account
explicitly in the calculations. This gives the
 one-gluon-exchange
(asymptotically leading) formula for the form factor in
terms of distribution amplitude $\phi(x)$ \cite{Brod}.
We recall, in passing,
 that the asymptotically leading contribution
predicts that the combination   $Q^2F_{\pi}(Q^2)$
is a constant.
Here the qualitative results of our  calculations
of the ``soft'' contribution, which is asymptotically suppressed
by power $1/Q^2$.

If we were started from the   wave function motivated   valence quark model
(\ref{BHL}) with mass parameter in it, than we could find the following
general behavior:
such a function gives a very reasonable
magnitude for $Q^2F_{\pi}(Q^2)$
in the intermediate region about few $GeV^2$
and it
starts to fall off very quickly right after that.
We expect that {\bf any} reasonable, well localized,
based on quark model wave function
with the scale $\sim \la\k\ra\sim m^2 $ leads to the similar
behavior.

 Currently, much more interesting for us is
the calculation of the soft contribution, based on the QCD motivated model
with Gaussian behavior and without mass term in it, see
(\ref{!}).
 The corresponding calculations show    the {\bf qualitative}
difference between  quark model and QCD- motivated wave functions.
Namely, the much slower fall  off at large $Q^2$
is observed for the $wf$ motivated by QCD.
The qualitative reason for that is the absence of the
mass term, see discussion after the formula (\ref{21}).
Precisely this term
was responsible for the very steep behavior in
 all previous calculations  based on a  quark model wave function.
 The declining of the form factor getting even slower
if one takes into account the property of the broadening
of  $wf$ in transverse direction, see\cite{Chibisov}
for details. This property corresponds to the
 strong fluctuations in the transverse
direction and quantitatively
is related to the large  parameter   $R$   ($\bullet{6}$).

One more qualitative remark.
The numerical magnitude   for the form factor strongly depends
on parameter $\la\xi^2\ra$. The soft contribution is getting bigger
when a wider (in longitudinal direction) wave function is used,
see \cite{Chibisov} for details.
 However, the most important observation that the QCD based
$wf$s {\bf simulate} the leading twist behavior, where
$Q^2F_{\pi}(Q^2)\simeq const.$ remains approximetely fulfilled.
 This   constant, however, itself is very sensitive
to the   numerical parameter $\la\xi^2\ra$. Whether this parameter
is large (as  QCD sum rules predict,   see ($\bullet 4$))
or small (close to the asymptotic value $\la\xi^2\ra=0.2$) is still a very
controversial issue. Our present attitude is that the true value,
as usual, is somewhere in the middle. In this case  the soft contribution
can be estimated as
$Q^2 F(Q^2)\sim 0.2\div 0.3.$, see \cite{Chibisov} for details.

The precise fitting of the pion form factor was not   among the goals of
these calculations. Rather, we wanted to demonstrate how the qualitative
properties of a nonperturbative $wf$, derived from the
QCD analysis might  significantly change its behavior.

Now we would like to extend our previous analysis
to the nucleon form factor. The starting point,
as before, is the fundamental constraints ($\bullet 1-\bullet 3$),
which being applied to the nucleon wave function imply
the Gaussian behavior with the specific argument (\ref{nucl}).
With these constraints in mind one can model the nucleon
$wf$ in the same way as we did for the pion. Having modeled
the nucleon wave function, one can calculate
a soft contribution to different
nucleon amplitudes. The corresponding analysis was carried out in the ref.
\cite{Kroll}. Here we quote some results from this paper.

  The most important qualitative result of these calculations
is similar to what we already observed previously in the $\pi$- meson case:
namely, the combination
$Q^4F^{nucl.}(Q^2)$ is almost constant in the extent region of $Q^2$
in spite of the fact that the corresponding
 ``soft'' contribution   naively  should
be decreasing function of  $Q^2$.
 The qualitative explanation
of this phenomenon is the same as before and is related to the
absence of the mass term in the QCD- motivated wave function.

The next   observation which was made in ref.\cite{Kroll}
  is related to the longitudinal distribution and can be formulated
as follows: A fit to the different experimental data leads
to a wave function which has  the same type of asymmetry
 which was  found previously
from the QCD sum rules. The asymmetry is however much more moderate
numerically than QCD sum rules indicate.
 We already mentioned earlier that the   similar conclusion
is likely to take place for the pion wave function   also.

 Therefore, the general moral, based on already completed calculations,
  can be formulated in the following way.
 There is a  standard   viewpoint for
the phenomenological success of the dimensional
counting rules: it is based on the
   prejudice  that the leading twist contribution
plays the main role in most cases. This outlook,
as we mentioned earlier, is based on the experimental data, where
the dimensional counting rules work very well.
We suggest here some different explanation for this
phenomenological success.
Our explanation of  the   slow falling off  of the soft contribution
with energy is due to the
specific properties of   nonperturbative $wf$.
 In particular, we argued that the absence of the
of the mass parameter in the corresponding $wf$ is the strict QCD constarint.
At the same time this property is responsible for
the  behavior mentioned above. Besides that,
we found a {\bf new} scale ($\sim 1GeV^2$) in the
 problem, in addition to the standard low energy parameter
$\la\k\ra\simeq 0.1 GeV^2$. Both these phenomena
lead to the  temporarily {\bf simulation} of the leading twist behavior
in the extent region of $ Q^2 $.
   We believe that this  is a new
  explanation of the phenomenological success of the dimensional
counting rules   at available, very modest energies.

Our next remark can be formulated as follows.
Exclusive, as well as inclusive amplitudes can be expressed
in terms   of the {\bf one and the same} particular hadron
$wf$. Therefore, if our explanation
(related to specific form of $\psi(\k,x)$) of a temporary simulation of
 the leading twist behavior  is considered as a reasonable one , then:

1. in the analysis of inclusive amplitudes
one may expect  the same effect ( it is our conjecture );

2. one may try to implement the intrinsic transverse
momentum dependence into the inclusive calculations.

In particular, one may try to use the following prescription
for the $\pi$ meson distribution function
(and anologously for nucleon, see (\ref{nucl})) at $x \Rightarrow 1$:
\be
\label{*}
G_{q/\pi}(x,Q^2)
\Rightarrow \{\int\frac{d^2k_{\perp}}{x(1-x)}\exp(- \frac{\k}{x(1-x)})
\} G_{q/\pi}(x,Q^2)
\ee

To support this conjecture, we would like to mention few inclusive processes
where the intrinsic
transverse distribution might be essential. First of all, it is
Drell-Yan amplitude $\pi+N\rightarrow\mu^-\mu^++X$
which is parametrized as follows (for
references and recent development see \cite{Khoze1}):
\be
\label{angle}
\frac{1}{\sigma}\frac{d\sigma}{d\Omega}\sim
1+\lambda\cos^2\theta+\mu\sin2\theta\cos\phi+\frac{\nu}{2}\sin^2\theta
\cos2\phi .
\ee
Here $\theta,\phi$ are angles defined in the muon pair rest frame and
$\lambda, \mu, \nu$ are coefficients.
In the naive parton model the coefficient are $\lambda=1,
\mu=\nu=0$. Experimental results do not support
this naive prediction. Recently, some improvements have been made
\cite{Khoze1}, but some problems are still remain. In particular,
the Lam-Tung sum rule \cite{Tung1}, $1-\lambda-2\nu=0$
is violated by experimental data and the improved model \cite{Khoze1}
still can not   explain the behavior
$1-\lambda-2\nu $ as a function of $Q_{\perp}^2$ (
$Q_{\perp}^2$ is the transverse
momentum
of the lepton pair).

Due to the fact that $Q^2$ is not large enough in this
experiment one may expect
that the intrinsic distribution (\ref{*}) might be essential.

One may find many examples like that where the standard
parton picture does not work well. We would like to mention
here the recent analysis \cite{Tung2}
of the direct photon production
($\pi+p\rightarrow\gamma + X$), with the result
that perturbative QCD can not explain the data. Some nonperturbative
broadening factor in transverse direction should be implemented.
One may hope that formula (\ref{*})
may improve the agreement with experiment.

Our last remark is the observation made in the recent
preprint \cite{Kroll} that the valence quark distribution functions
$u^p(x)$ and $d^p(x)$ at large $x$  can be fairly described
in terms of  the nonperturbative nucleon $wf$. Two  properties
of the  $wf$ are important to provide such a fit:
The absence of the mass term in the formula (\ref{nucl}) (this leads to the correct
power behavior at $x\rightarrow 1$) and a moderate asymmetry
in the longitudinal direction ( this provides an observed ratio for
the $\frac{u^p(x)}{d^p(x)}\simeq 5 ~at~~ x\rightarrow 1$ in the contrast with
the asymptotic formula prediction which gives value of 2 for the same ratio).
\section{Conclusion}
We believe that the most important result of the present analysis
is the observation that due to the
specific properties of   $\psi(\k, x)$, the ``soft'' contribution
can  temporarily {\bf simulate} the leading twist behavior
in the extent region of $ Q^2:~~3 GeV^2\leq Q^2\leq 40 GeV^2 $.
 Such a mechanism, if it is correct, would be
 an explanation of the phenomenological success of the dimensional
counting rules   at available, very modest energies
for many different processes.

\end{document}